# Optical radiation force (per–length) on an electrically conducting elliptical cylinder having a smooth or ribbed surface

F.G. MITRI[*]

*Santa Fe, NM 87508, USA*
*\*f.g.mitri@ieee.org*

**Abstract:** The aim of this work is to develop a formal semi-analytical model using the modal expansion method in cylindrical coordinates to calculate the optical/electromagnetic (EM) radiation force-per-length experienced by an infinitely long electrically-conducting elliptical cylinder having a smooth or wavy/corrugated surface in EM plane progressive waves with different polarizations. In this analysis, one of the semi-axes of the elliptical cylinder coincides with the direction of the incident field. Initially, the modal matching method is used to determine the scattering coefficients by imposing appropriate boundary conditions and solving numerically a linear system of equations by matrix inversion. In this method, standard cylindrical (Bessel and Hankel) wave functions are used. Subsequently, simplified expressions leading to exact series expansions for the optical/EM radiation forces assuming either electric (TM) of magnetic (TE) plane wave incidences are provided without any approximations, in addition to integral equations demonstrating the direct relationship of the radiation force with the square of the scattered field magnitude. An important application of these integral equations concerns the accurate determination of the radiation force from the measurement of the scattered field by any 2D non-absorptive object of arbitrary shape in plane waves. Numerical computations for the non-dimensional radiation force function are performed for electrically conducting elliptic and circular cylinders having a smooth or ribbed/corrugated surface. Adequate convergence plots confirm the validity and correctness of the method to evaluate the radiation force with no limitation to a particular frequency range (i.e. Rayleigh, Mie or geometrical optics regimes). Particular emphases are given on the aspect ratio, the non-dimensional size of the cylinder, the corrugation characteristic of its surface, and the polarization of the incident field. The results are particularly relevant in optical tweezers and other related applications in fluid dynamics, where the shape and stability of a cylindrical drop stressed by a uniform external electric/magnetic field are altered. Furthermore, a direct analogy with the acoustical counterpart is noted and discussed.

## 1. Introduction

Cylinders with non-circular geometrical cross-sections are encountered in various applications in electromagnetic (EM)/optical scattering [1-5], particle characterization [6] and manipulation, and computer visualization and graphics [7] to name a few areas. They received significant interest from the standpoint of optical scattering theory in elliptical coordinates [8, 9], cylindrical coordinates [10], and vector complex ray modeling [11]. Moreover, investigations on the optical radiation forces [12] and torques [13] have been performed, which were based solely on the brute force numerical integration of Maxwell's radiation stress over the surface of the cylinder.

The EM radiation force and torque are intrinsically connected with the scattering of the incident illuminating field by the cylinder due to the transfer of linear and angular momenta. Although the earlier works [12, 13] considered some developments for the optical force and torque on an elliptic cylinder with a smooth surface using the brute force numerical integration method or the boundary element method [14], an improved formalism showing explicitly the dependence of the radiation force on the scattering coefficients of the corrugated non-circular cylinder is advantageous from a physical understanding standpoint.

The aim of this work is therefore directed toward developing a semi-analytical formalism based on the partial wave series expansion method in cylindrical coordinates [10] without any approximations. Another method may consider a wave-basis using infinite series expansions in elliptical Mathieu functions [15-21], however, the generation and computation of the elliptical functions may not be a straightforward task in numerical implementation.

To alleviate potential impediments with Mathieu functions, the formalism based on the standard cylindrical Bessel and Hankel wave functions introduced previously in the acoustical context in 2D [22-26] is extended here to the case of an electrically-conducting elliptical/circular cylinder with either a smooth or corrugated surface. The formalism is applicable to any range of frequencies such that either the long- or short-wavelength with respect to the size of the cylinder can be examined rigorously. Notice that apart from their practical importance, in most cases two-dimensional scattering and radiation force problems provide rigorous, convenient and efficient ways of displaying emergent physical phenomena without involving the mathematical/algebraic manipulations encountered in solving three-dimensional problems [27-33], and this work should assist in further developing analytical radiation force models for spheroidal [27, 30] and ellipsoidal particles with smooth or corrugated surfaces.

## 2. Electric and magnetic wave incidence cases (known also as TM and TE polarizations, respectively) and analysis of the scattering

Initially, the case of incident electrical waves are considered, meaning that the electric field polarization is along the $z$-axis perpendicular to the polar plane $(r, \theta)$ as shown in Fig. 1 (known also as TM polarization; i.e., $H_z^{inc} = 0$, where $H_z^{inc}$ is the axial component of the magnetic field).

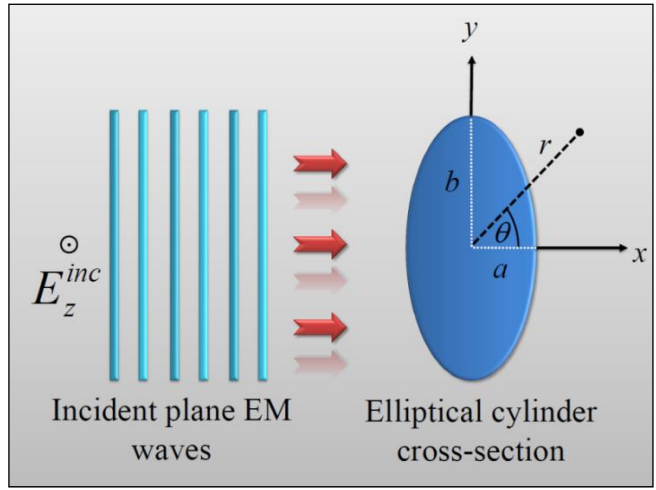

Fig. 1. A sketch describing the interaction of an incident electric field polarized along the axial $z$-direction perpendicular to the plane of the figure (known also as TM polarization) with a perfect electrically conducting elliptical smooth cross-section in a homogeneous medium. The semi-axes of the ellipse are denoted by $a$ and $b$, respectively. The cylindrical coordinate system $(r, \theta, z)$ is referenced to the center of the elliptical cylinder.

Consider a plane progressive wave propagating in a homogeneous non-absorptive non-magnetic medium, and incident upon an elliptical cylinder where one of the semi-axes lies on the axis of wave propagation *x* of the incident waves (i.e., on-axis configuration).

The axial component of the incident electric field vector can be expressed in terms of a series expansions in cylindrical coordinates ($r$, $\theta$, $z$) such that,

$$E_z^{inc} = E_0 e^{-i\omega t} \sum_n i^n J_n(kr) e^{in\theta}, \qquad (1)$$

where $E_0$ is the amplitude, the summation $\sum_n = \sum_{n=-\infty}^{+\infty}$, $J_n(\cdot)$ is the cylindrical Bessel function of first kind of order $n$, $\theta$ is the polar angle, and $k$ is the wave number of the incident radiation.

The scattered electric field component is also expressed as a series expansion as,

$$E_z^{sca} = E_0 e^{-i\omega t} \sum_n i^n C_n^{TM} H_n^{(1)}(kr) e^{in\theta}, \qquad (2)$$

where $C_n^{TM}$ are the scattering coefficient to be determined by applying the necessary boundary condition for a perfectly conducting cylinder, and $H_n^{(1)}(.)$ is the cylindrical Hankel function of the first kind, describing the propagation of outgoing waves.

The cylinder boundary corresponds to an ellipse having a shape function $A_\theta$. For a corrugated elliptical surface, its expression is given by [34],

$$A_\theta = \left[(\cos\theta/a)^2 + (\sin\theta/b)^2\right]^{-1/2} + d\cos(\ell\theta), \qquad (3)$$

where $a$ and $b$ are the semi-axes of the ellipse, $d$ is the amplitude of the surface roughness, and $\ell$ is an integer number which determines the periodicity of the surface corrugations.

Application of the boundary condition for a perfect electrically-conducting cylinder requires that the *tangential component* of the total (i.e., incident + scattered) electric field vector vanishes at $r = A_\theta$, such that,

$$\left.(\mathbf{n} \times \mathbf{E}) \cdot \mathbf{e}_\theta\right|_{r=A_\theta} = 0, \qquad (4)$$

where the normal vector **n** is expressed as,

$$\mathbf{n} = \mathbf{e}_r - \left(\frac{1}{A_\theta}\right)\frac{dA_\theta}{d\theta}\mathbf{e}_\theta, \qquad (5)$$

with $\mathbf{e}_r$ and $\mathbf{e}_\theta$ denoting the outward unit vectors along the radial and tangential directions, respectively.

Substituting Eqs.(1) and (2) into Eqs.(4) using (5) leads to a system of linear equations,

$$\sum_n \mathrm{i}^n \left[ \Gamma_n^{\mathrm{TM}}(\theta) + C_n^{\mathrm{TM}} \Upsilon_n^{\mathrm{TM}}(\theta) \right] = 0, \qquad (6)$$

where the structural functions $\Gamma_n^{\mathrm{TM}}(\theta)$ and $\Upsilon_n^{\mathrm{TM}}(\theta)$ are expressed, respectively, as

$$\begin{Bmatrix} \Gamma_n^{\mathrm{TM}}(\theta) \\ \Upsilon_n^{\mathrm{TM}}(\theta) \end{Bmatrix} = \mathrm{e}^{\mathrm{i} n \theta} \begin{Bmatrix} J_n(kA_\theta) \\ H_n^{(1)}(kA_\theta) \end{Bmatrix}. \qquad (7)$$

The angular dependency in the structural functions should be eliminated in order to solve the system of linear equations for each partial-wave *n* mode. Therefore, Eq.(6) is equated to a Fourier series as,

$$\sum_n \mathrm{i}^n \left[ \Gamma_n^{\mathrm{TM}}(\theta) + C_n^{\mathrm{TM}} \Upsilon_n^{\mathrm{TM}}(\theta) \right] = \sum_n \left[ \psi_n^{\mathrm{TM}} + C_n^{\mathrm{TM}} \Omega_n^{\mathrm{TM}} \right] \mathrm{e}^{\mathrm{i} n \theta} \qquad (8)$$
$$= 0.$$

The coefficients $\psi_n^{\mathrm{TM}}$ and $\Omega_n^{\mathrm{TM}}$ are now independent of the angle $\theta$, and they are determined after applying to Eq.(8) the following orthogonality condition,

$$\int_0^{2\pi} \mathrm{e}^{\mathrm{i}(n-m)\theta} d\theta = 2\pi \delta_{n,m}, \qquad (9)$$

where $\delta_{i,j}$ is the Kronecker delta function. This procedure leads to a new system of linear equations expressed as,

$$\sum_m \left[ \psi_{n,m}^{\mathrm{TM}} + C_n^{\mathrm{TM}} \Omega_{n,m}^{\mathrm{TM}} \right] = 0, \qquad (10)$$

where $\sum_m = \sum_{m=-\infty}^{+\infty}$, and,

$$\begin{Bmatrix} \psi_{n,m}^{\mathrm{TM}} \\ \Omega_{n,m}^{\mathrm{TM}} \end{Bmatrix} = \frac{1}{2\pi} \sum_n \mathrm{i}^n \int_0^{2\pi} \begin{Bmatrix} \Gamma_n^{\mathrm{TM}}(\theta) \\ \Upsilon_n^{\mathrm{TM}}(\theta) \end{Bmatrix} \mathrm{e}^{-\mathrm{i} m \theta} d\theta. \qquad (11)$$

For the magnetic wave incidence case (known also as TE polarization; i.e., $E_z^{inc} = 0$), application of the boundary condition requires that the tangential component of the total (i.e., incident + scattered) electric field vector vanishes at $r = A_\theta$, such that,

$$(\mathbf{n} \times \mathbf{E}) \cdot \mathbf{e}_z \big|_{r=A_\theta} = 0, \qquad (12)$$

with $\mathbf{e}_z$ denoting the outward unit vector along the axial direction. This leads to a new system of equations that must be solved, as is given as

$$\sum_m \left[ \psi_{n,m}^{\mathrm{TE}} + C_n^{\mathrm{TE}} \Omega_{n,m}^{\mathrm{TE}} \right] = 0, \qquad (13)$$

where,

$$\left\{\begin{array}{l}\psi_{n,m}^{\text{TE}}\\ \Omega_{n,m}^{\text{TE}}\end{array}\right\}=\frac{1}{2\pi}\sum_{n}\mathrm{i}^{n}\int_{0}^{2\pi}\left\{\begin{array}{l}\Gamma_{n}^{\text{TE}}(\theta)\\ \Upsilon_{n}^{\text{TE}}(\theta)\end{array}\right\}\mathrm{e}^{-\mathrm{i}m\theta}d\theta, \quad (14)$$

and the structural functions $\Gamma_{n}^{\text{TE}}(\theta)$ and $\Upsilon_{n}^{\text{TE}}(\theta)$ are expressed, respectively, as

$$\left\{\begin{array}{l}\Gamma_{n}^{\text{TE}}(\theta)\\ \Upsilon_{n}^{\text{TE}}(\theta)\end{array}\right\}=\mathrm{e}^{\mathrm{i}n\theta}\left[k\left\{\begin{array}{l}J_{n}^{'}(kA_{\theta})\\ H_{n}^{(1)'}(kA_{\theta})\end{array}\right\} - \mathrm{i}\left(\frac{n}{A_{\theta}^{2}}\right)\frac{dA_{\theta}}{d\theta}\left\{\begin{array}{l}J_{n}(kA_{\theta})\\ H_{n}^{(1)}(kA_{\theta})\end{array}\right\}\right]. \quad (15)$$

The primes in Eq.(15) denote a derivative with respect to the argument of the cylindrical wave functions.

Away from the elliptic cylinder in the far-field, the cylindrical Hankel function of the first kind is approximated by its asymptotic limit $H_{n}^{(1)}(kr) \underset{kr\to\infty}{\approx} \mathrm{i}^{-n}\sqrt{2/(\mathrm{i}\pi kr)}\,\mathrm{e}^{\mathrm{i}kr}$. Therefore, it is convenient to define non-dimensional (steady-state, i.e., time-independent) far-field scattering form functions as,

$$\left\{\begin{array}{l}f_{\infty}^{\text{TM}}(k,\theta)\\ f_{\infty}^{\text{TE}}(k,\theta)\end{array}\right\}=\sqrt{(a_{\text{eff}}/a)}\sqrt{(2r/a)}\left\{\begin{array}{l}E_{z}^{sca}/E_{0}\\ H_{z}^{sca}/H_{0}\end{array}\right\}\mathrm{e}^{-\mathrm{i}(kr-\omega t)}$$
$$=\frac{2}{\sqrt{\mathrm{i}\pi ka}}\sqrt{(a_{\text{eff}}/a)}\sum_{n}\left\{\begin{array}{l}C_{n}^{\text{TM}}\\ C_{n}^{\text{TE}}\end{array}\right\}\mathrm{e}^{\mathrm{i}n\theta}, \quad (16)$$

where $H_{z}^{sca}$ is the axial component of the scattered magnetic field vector, $H_{0}$ is the amplitude of the magnetic field, and $a_{\text{eff}}$ is an effective radius defined as $a_{\text{eff}}=\sqrt{(a^{2}+b^{2})/2}$. The factor $\sqrt{(a_{\text{eff}}/a)}$ equals unity for a cylinder with circular cross-section.

### 3. Simplified radiation force expressions for the electric (TM) and magnetic (TE) wave incidence cases

Evaluation of the EM radiation force stems from an analysis of the far-field scattering [35], which does not introduce any approximation in a homogeneous non-absorptive medium as recognized in various investigations [36-39].

As such, the general expression for the EM radiation force vector in 2D given earlier [40], and applied to the electric wave incidence case is written as,

$$\left\langle \mathbf{F}^{TM} \right\rangle_{kr\to\infty} = \frac{1}{2c}\Re\left\{ \iint_S \left[ \begin{array}{c} E_{z,\infty}^{inc*}H_{\theta,\infty}^{sca} + E_{z,\infty}^{sca}H_{\theta,\infty}^{inc*} \\ + E_{z,\infty}^{sca*}H_{\theta,\infty}^{sca} \end{array} \right] d\mathbf{S} \right\}, \quad (17)$$

where the integration of the EM linear momentum is performed over a surface $S$ enclosing the object, $\Re\{\cdot\}$ denotes the real part of a complex number, $c$ is the speed of light, and the subscript $\infty$ indicates that the far-field limits of the electric and magnetic field components are used. Subsequently, Eq.(17) can be expressed in a simplified form as

$$\left\langle \mathbf{F}^{TM} \right\rangle_{kr\to\infty} = \frac{\sqrt{\varepsilon}}{2c} \iint_S \Re\{E_{is}\} d\mathbf{S}, \quad (18)$$

where $E_{is} = E_{z,\infty}^{sca*}\left[ (i/k)\partial_r E_{z,\infty}^{inc} - E_{z,\infty}^{inc} - E_{z,\infty}^{sca} \right]$, the differential operator is $\partial_r = \partial/\partial r$, and $k$ is the wavenumber in the medium of wave propagation.

Assuming now magnetic wave (TE) incidence, the EM radiation force vector is expressed as [40],

$$\left\langle \mathbf{F}^{TE} \right\rangle_{kr\to\infty} = -\frac{1}{2c}\Re\left\{ \iint_S \left[ \begin{array}{c} E_{\theta,\infty}^{inc*}H_{z,\infty}^{sca} + E_{\theta,\infty}^{sca}H_{z,\infty}^{inc*} \\ + E_{\theta,\infty}^{sca}H_{z,\infty}^{sca*} \end{array} \right] d\mathbf{S} \right\}. \quad (19)$$

With further simplifications, Eq.(19) is rewritten in a simplified form as,

$$\left\langle \mathbf{F}^{TE} \right\rangle_{kr\to\infty} = \frac{1}{2c\sqrt{\varepsilon}} \iint_S \Re\{H_{is}\} d\mathbf{S}, \quad (20)$$

where $H_{is} = H_{z,\infty}^{sca*}\left[ (i/k)\partial_r H_{z,\infty}^{inc} - H_{z,\infty}^{inc} - H_{z,\infty}^{sca} \right]$.

Eqs.(18) and (20) correspond to the simplified expressions for the radiation force vector in each polarization scheme. The differential surface vector is expressed as $d\mathbf{S} = Lrd\theta\mathbf{e}_r$, where $L$ is the length of a cylindrical surface enclosing the particle, and the outward normal unit vector in the radial direction is $\mathbf{e}_r = \cos\theta\mathbf{e}_x + \sin\theta\mathbf{e}_y$, where $\mathbf{e}_x$ and $\mathbf{e}_y$ are the unit vectors in the Cartesian coordinates system.

### 4. Integral expressions for the longitudinal radiation force (i.e. along the *x*-direction) assuming plane wave incidence on a non-absorptive object

Assuming first an optical/EM plane progressive wave field with TM polarization propagating along the *x*-direction as shown in Fig. 1, the expression for the longitudinal radiation force per-length $F_x^{TM}/L = \left\langle \mathbf{F}^{TM} \right\rangle \cdot \mathbf{e}_x /L$, can be obtained from Eq.(18) such that,

$$\frac{F_x^{TM}}{L} = -\frac{\sqrt{\varepsilon}}{2c} r \int_0^{2\pi} \Re\left\{E_z^{inc} E_z^{sca*}(1+\cos\theta) + \left|E_z^{sca}\right|^2 \cos\theta\right\} d\theta. \quad (21)$$

Further simplification of Eq.(21) can be attained by noticing the following property for a non-absorptive object,

$$\int_0^{2\pi} \Re\left\{E_z^{inc} E_z^{sca*}(1+\cos\theta)\right\} d\theta = -\int_0^{2\pi} \left|E_z^{sca}\right|^2 d\theta. \quad (22)$$

Eq.(22) stems from the statement of energy conservation applied to scattering [41-43], affirming that for a non-absorptive object, both the scattering and extinction EM powers are equal. Thus, Eq.(21) gives the longitudinal optical/EM radiation force component as,

$$F_x^{TM} = \frac{\sqrt{\varepsilon}}{2c} \int_0^{2\pi} \left|E_z^{sca}\right|^2 (1-\cos\theta) dS. \quad (23)$$

As discussed previously [40], a dimensionless radiation force function for the elliptical cylinder can be defined such that $Y_p^{TM} = cF_x^{TM}/(2aLI_0)$, where $I_0 = \sqrt{\varepsilon}|E_0|^2/2$. Based on Eq.(23), it is expressed as,

$$Y_p^{TM} = (a/a_{eff}) \frac{1}{4} \int_0^{2\pi} \left|f_\infty^{TM}(k,\theta)\right|^2 (1-\cos\theta) d\theta, \quad (24)$$

where the far-field scattering form function $f_\infty^{TM}(k,\theta)$ is given by Eq.(16).

Considering now the case of a plane magnetic wave incidence with TE polarization, equivalent expressions for $F_x^{TE}$ and $Y_p^{TE}$ are obtained, respectively, as,

$$F_x^{TE} = \frac{1}{2c\sqrt{\varepsilon}} \int_0^{2\pi} \left|H_z^{sca}\right|^2 (1-\cos\theta) dS, \quad (25)$$

and

$$Y_p^{TE} = (a/a_{eff}) \frac{1}{4} \int_0^{2\pi} \left|f_\infty^{TE}(k,\theta)\right|^2 (1-\cos\theta) d\theta. \quad (26)$$

As an end result, the substitution of the form functions given by Eq.(16) into Eqs.(24) and (26) and manipulation of the angular integral leads to the *exact* series expansions for the longitudinal dimensionless radiation force functions as,

$$Y_p^{\{TM,TE\}} = -\frac{1}{ka} \Re\left\{\sum_n C_n^{\{TM,TE\}*}\left(C_{n+1}^{\{TM,TE\}} + C_{n-1}^{\{TM,TE\}} + 2\right)\right\}. \quad (27)$$

Eqs.(23)-(26) show important results where the immediate connection of the longitudinal force component (or the longitudinal dimensionless force function) with the axial component of the electric (or magnetic) field is established. The significance of these results is evaluated from these equations such that an estimation of the force on a *non-absorptive object in plane waves* can be accomplished by measuring the scattered electric (or magnetic) field experimentally around the object at each polar angle $0 \leq \theta \leq 2\pi$, and integrate the square of its magnitude according to Eqs.(23)-(26). Moreover, the forms of Eqs.(24) and (26) indicate that $Y_p^{\{TM,TE\}}$ is $\geq 0$. Equivalent expressions in the acoustical context for circular cylinders have been previously obtained as well [24, 44, 45].

Notice that the radiation force function expression given by Eq.(27) can be reformulated as a function of phase shifts based on the fundamental analysis of the wave equation in electromagnetic/acoustic scattering (pp. 1376 – 1382 in [46]), the nuclear resonance scattering theory (see Section 3.12, pp. 148 – 152 in [47]) or quantum physics (see Section 3.3, pp. 159 – 200 in [48]). In essence, the scattering coefficients $C_n^{\{TM,TE\}}$ in Eq.(27) can be rewritten in terms of the phase shifts (usually denoted by $\delta_n^{\{TM,TE\}}$) as $C_n^{\{TM,TE\}} = i \sin \delta_n^{\{TM,TE\}} e^{i\delta_n^{\{TM,TE\}}}$. The use of phase shifts in radiation force theory is a convenient way to parameterize the scattering and analyze the contribution of resonances and other effects in any scattering regime (i.e., Rayleigh, Mie and geometrical optical). For a lossless material, the phase shifts $\delta_n^{\{TM,TE\}}$ are real numbers, whereas for an absorptive (or radiating) particle, those coefficients become complex where the imaginary part represents the rate of absorption [41] (for a passive object) or amplification [49] (for an active emitting source object). In terms of the phase shifts, Eq.(27) becomes,

$$Y_p^{\{TM,TE\}} = -\frac{1}{ka} \Re \left\{ \sum_n \sin \delta_n^{\{TM,TE\}*} e^{-i\delta_n^{\{TM,TE\}*}} \right. \\
\left. \times \left( \sin \delta_{n+1}^{\{TM,TE\}} e^{i\delta_{n+1}^{\{TM,TE\}}} + \sin \delta_{n-1}^{\{TM,TE\}} e^{i\delta_{n-1}^{\{TM,TE\}}} + 2 \right) \right\}. \quad (28)$$

## 5. Computational results and discussions

The analysis is concentrated on computing the dimensionless radiation force functions for electrically-conducting elliptical cylinders in TM and TE polarized plane waves as given by Eq.(27). Initially, a smooth elliptical cylinder is considered such that $d = 0$ in Eq.(3). As seen from Eq.(27), $Y_p^{\{TM,TE\}}$ depends on the scattering coefficients $C_n^{\{TM,TE\}}$ that must be computed first.

The initial task consists of computing those coefficients given by Eqs.(10) and (13) using matrix inversion procedures [22, 24, 25]. The related integrals given by Eqs.(11) and (14) are computed using Simpson's rule for numerical integration with a sampling of 650 points of the surface and 50 multipoles leading to negligible numerical errors (see subsequently the convergence plots). Notice that the maximum truncation limit in the series must be increased if higher frequencies (i.e. $kb > 5$) are considered to ensure adequate convergence.

Validation and verification of the formal solutions are performed and presented in panels (a) and (b) of Fig. 2 for the TM and TE polarization cases, respectively, considering an elliptical cylinder with a smooth surface. These tests of convergence required evaluating $Y_p^{\{TM,TE\}}$ versus the maximum truncation limit $n_{max}$ varying in the range $0 \leq n_{max} \leq 50$ for $kb$ = 5. These plots show that the proper convergence to the stable/steady solution is the fastest for a circular cylinder (i.e., $a/b = 1$, corresponding to the plot with the triangle pointing downwardly) requiring the least number of partial waves ~ 7 to ensure the correctness of the results. Furthermore, the convergence is faster for an ellipse with an aspect ratio $a/b = 0.5$, while for an ellipse with an aspect ratio $a/b = 2$, thirty (i.e., $n_{max} = 30$) terms are needed as a minimum to ensure adequate convergence with minimal truncation error. In essence, the plots displayed in panels (a) and (b) attest of the correctness of the solution.

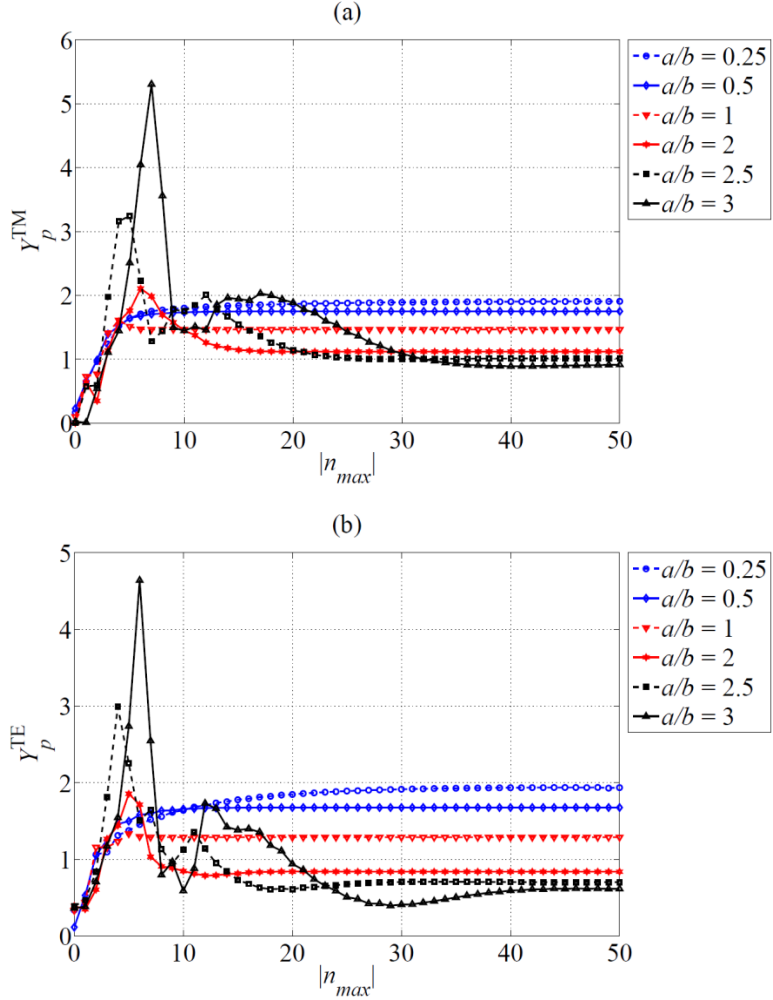

Fig. 2. Panel (a) displays the convergence plots for the dimensionless radiation force function versus the maximum truncation order $|n_{max}|$ assuming TM polarized plane waves at $kb = 5$ for various aspect ratios of the elliptical cylinder with a smooth surface. The circular cylinder case corresponds to the plot having $a/b = 1$. Panel (b) displays similar convergence plots assuming TE polarized plane progressive waves.

Next, computations for $Y_p^{\{TM,TE\}}$ are performed in the range $0 < kb \leq 5$ with emphasis on polarization and the aspect ratio of the elliptical cylinder centered on the axis of propagation of the incident field (i.e., on-axis configuration).

Panel (a) of Fig. 3 shows the results for $Y_p^{TM}$ for a Rayleigh elliptical cylinder where for $a/b = 0.25$, the radiation force function is the smallest for $kb < 0.7$, while for $a/b = 3$, it is the largest. Around $kb \sim 0.7$, panel (a) shows that the plots for $Y_p^{TM}$ exhibit comparable values for the chosen aspect ratios. Beyond that limit (i.e., as $kb$ increases $> 0.7$), the radiation force function for $a/b = 0.25$ becomes the largest, while that related to $a/b = 3$ turns out to be the smallest as shown in the plots of panel (b).

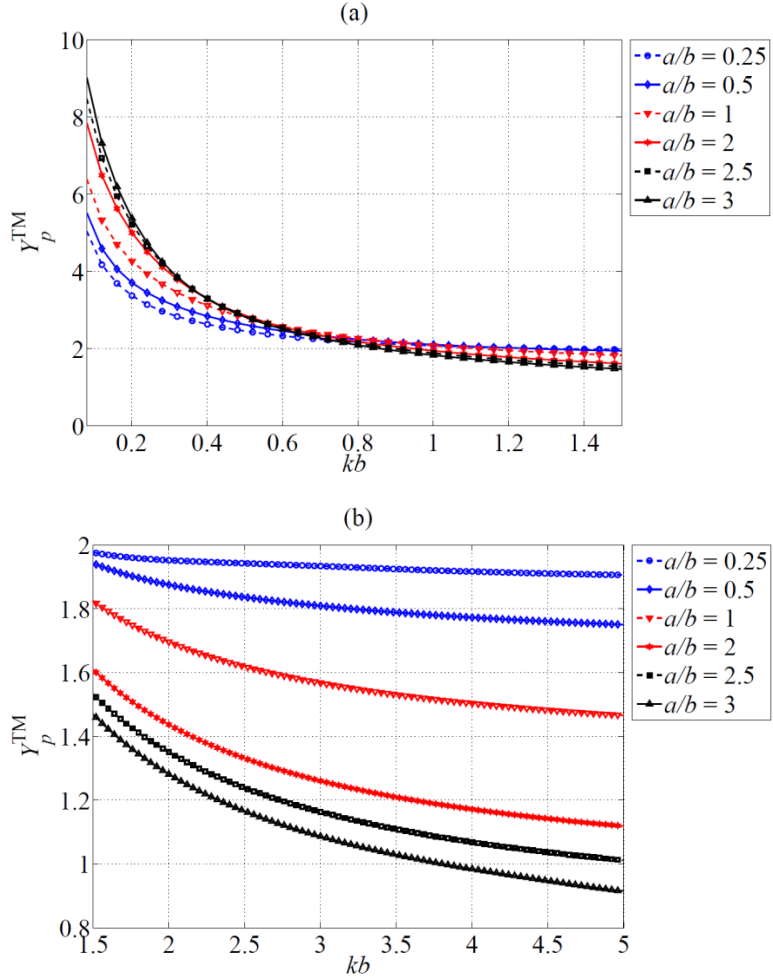

Fig. 3. Panel (a) displays the plots for the dimensionless radiation force function assuming TM polarization of the incident plane waves for Rayleigh electrically conducting elliptical cylinders of various aspect ratios $a/b$ versus $kb$. Panel (b) corresponds to the same plots but in a larger $kb$-range.

The effect of changing the polarization to TE on the radiation force function is also investigated for the elliptic cylinder with a smooth surface, and the corresponding results are displayed in panels (a) and (b) of Fig. 4 for a Rayleigh and Mie elliptical cylinders, respectively. Visual comparison of the plots with those of Fig. 3 shows the clear differences from the TM polarization case. Nonetheless, the radiation force function behavior for the

Rayleigh elliptical cylinder case shown in panel (a) of Fig. 4 is quite similar to that of panel (a) of Fig. 3; i.e., for $kb < 0.4$, the radiation force function is the smallest for $a/b = 0.25$, while for $a/b = 3$, it is the largest. The opposite effect occurs as $kb$ increases $> 0.4$ as shown in panel (b) of Fig. 4.

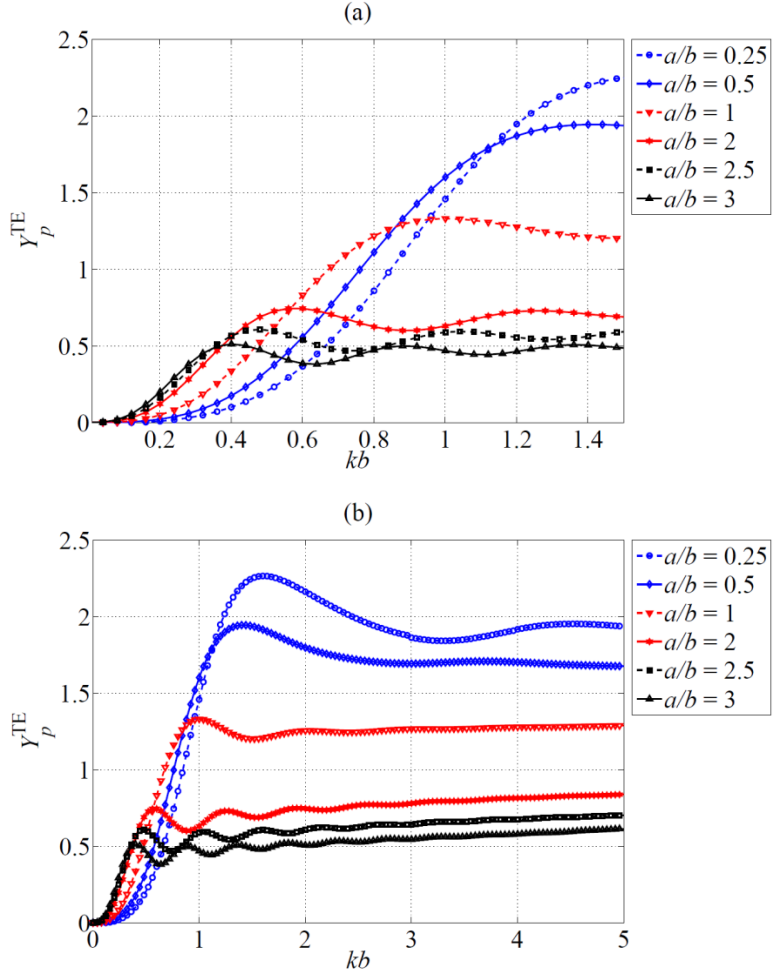

Fig. 4. The same as in Fig. 3, but the incident plane wave field is TE polarized.

Corrugated cylinders with a wavy/ribbed surface are now considered as shown in panels (a)-(c) of Fig. 5. In all these cases, a deformation parameter $d/a = 0.1$ is considered with a periodicity $\ell = 10$ while the aspect ratio is varied.

Convergence plots are computed for the radiation force function assuming TE polarization, and the results are displayed in Fig. 6 for different aspect ratios $a/b = 0.5$, 1 and 1.5, respectively. Convergence is the fastest for the circular corrugated cylinder (i.e., $a/b = 1$) requiring the least number of partial waves $\sim n_{max} = 18$, while for the elongated corrugated case for $a/b = 1.5$, $n_{max} = 50$ is required.

The plots in Fig. 7 correspond to the radiation force function in the TE polarization case for corrugated elliptical and circular cylinders versus $kb$. These plots also show that $Y_p^{TE}$ for an elliptical corrugated cylinder with a defined aspect ratio behaves differently in the Rayleigh versus the Mie regimes. For $kb < 0.4$, the corrugated cylinder having $a/b = 1.5$ experiences the largest force, while the one having $a/b = 0.5$ is subjected to the smallest. This same behavior also occurs as $kb$ increases $> 3$, where this increase does not occur for the elliptical cylinders with a smooth surface. Also the corrugated circular cylinder result (i.e., $a/b = 1$) shows that the radiation force function is larger than that of the elliptical cylinder case with $a/b = 0.5$ for $kb > 3.5$.

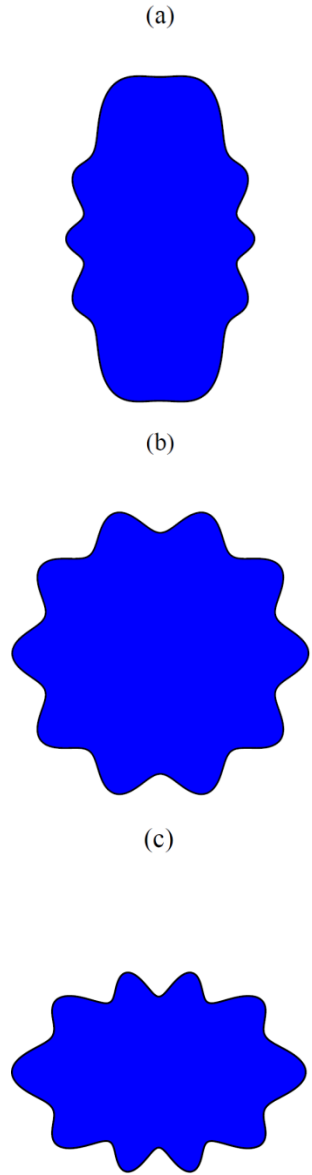

Fig. 5. Panel (a) displays the plot for the surface shape function of a corrugated elliptical cylinder having an aspect ratio $a/b = 0.5$ and $\ell = 10$. Panel (b) corresponds to the case where $a/b = 1$, while panel (c) is for an aspect ratio $a/b = 1.5$. In all the panels, $d/a = 0.1$.

To further highlight the effect of the surface corrugations on the radiation force function $Y_p^{TE}$, additional plots with and without corrugations for the same elliptical/circular cylinder has been performed and displayed in panels (a)-(c) of Fig. 8 for $a/b$ = 0.5, 1 and 1.5, respectively. Those plots clearly show that for the case corresponding to panel (a) of Fig. 5, the effects of corrugations are negligible for $kb < 4$, but should be considered when $kb > 4$. This is not the case for the corrugated circular cylinder case shown in panel (b) as $kb > 0.5$, as well as the one displayed in panel (c). These plots unequivocally show that the corrugations can enhance the radiation force on the elliptical/circular cylinder by several orders of magnitude depending on $kb$.

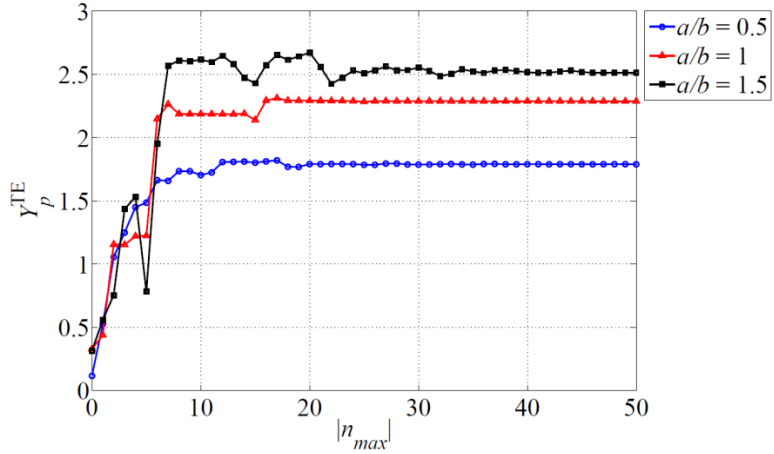

Fig. 6. Convergence plots for the dimensionless radiation force function versus the maximum truncation order $n_{max}$ assuming TE polarized plane waves at $kb = 5$ for various aspect ratios of the elliptical cylinder with a corrugated surface having $d/a = 0.1$ and $\ell = 10$.

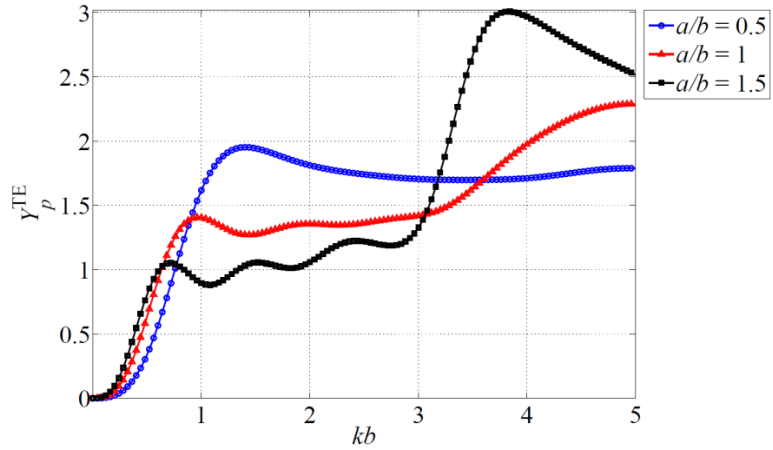

Fig. 7. The plots for the dimensionless radiation force function assuming TE polarization of the incident plane waves for electrically conducting elliptical cylinders with corrugated surfaces [as shown in Fig. 5] of various aspect ratios $a/b$ versus $kb$.

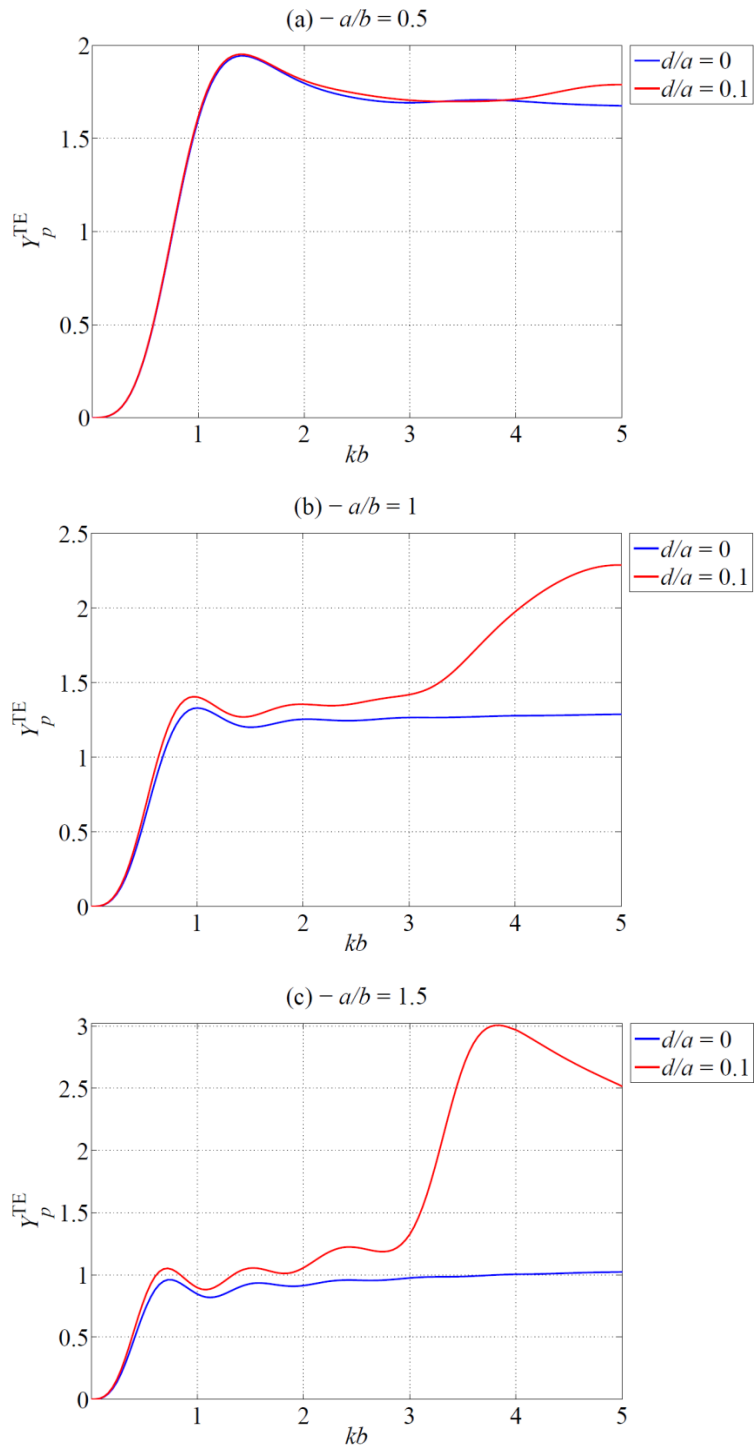

Fig. 8. Panel (a) shows a comparison of the plots for the dimensionless radiation force function assuming TE polarization of the incident plane waves for an electrically conducting elliptical cylinder having an aspect ratio $a/b = 0.5$ without ($d/a = 0$) and with corrugations ($d/a = 0.1$). Panels (b) and (c) display similar plots but for aspect ratios $a/b = 1$ and 1.5, respectively.

## 6. Conclusion and perspectives

This analysis presented formal exact series expansions without any approximations in cylindrical coordinates for the radiation force per-length and its dimensionless function assuming TM and TE polarizations of an incident plane progressive wave field. The method developed here utilizes standard cylindrical Bessel and Hankel functions as the basis functions, and allows computing accurately the scattering coefficients of an elliptic electrically-conducting cylinder having either a smooth or corrugated/ribbed surface without limitation to a particular frequency range such that the Rayleigh, Mie and the geometrical optics limits can be all considered if desired. Integral equations showing the direct connections of the radiation force functions with the square of the scattering form functions in the far-field from a non-absorptive scatterer (applicable for plane progressive waves) are developed and discussed. Moreover, numerical computations for the radiation force functions illustrate the analysis with particular emphasis on the aspect ratio of the elliptical cylinder, the non-dimensional size parameter $kb$, as well as the corrugation and periodicity parameters. Notice that the benchmark semi-analytical solutions developed here could be used to validate results obtained by strictly numerical methods, such as the finite/boundary element methods.

Although the present analysis have dealt with a perfectly conducting elliptical cylinder with a smooth or a corrugated surface, the extension to the case of a permeable dielectric particle is straightforward. It must be emphasized, however, that the boundary conditions given by Eqs.(4) and (12) for the TM and TE polarizations, respectively, must be substituted by an appropriate set of equations such that the tangential components of the total electric and magnetic fields are *continuous* at the boundary of the particle (i.e., at $r = A_\theta$). Consequently, new coupled systems of equations can be obtained for each type of polarization, respectively, which must be solved by matrix inversion in order to compute the scattering coefficients $C_n^{\{TM,TE\}}$ of the dielectric particle of arbitrary shape. Then, the radiation force functions given in Section 4 can be evaluated accordingly.

Potential applications of the present analysis are in fluid dynamics to study the deformation [50, 51] and stability of a liquid bridge in electric fields [52, 53], opto-fluidic devices, and particle manipulation to name a few examples. Moreover, the scope of the present analysis can be extended to evaluate the radiation force on multiple corrugated particles [54] based on the recently developed formalisms [55, 56], and in the presence of a boundary [49, 57, 58] or a corner-space [59] located nearby the particle. The extension of the analysis to consider other wavefronts (differing from plane waves) such as non-paraxial Airy [60-66] and Gaussian [61, 67-70] light sheets is also possible.

Regarding the acoustical analog, it is important to note that the series expressions and corresponding results presented here are directly applicable to the case of soft (TM) or rigid (TE) elliptical cylinders with smooth or corrugated surfaces. Recent advances in the acoustical context for non-circular geometries [22-25] have considered such cases in plane waves [23-25] and non-paraxial Gaussian acoustical-sheets [22].